\documentstyle[prb,aps]{revtex}
\begin{document}
\title{Interfacial adsorption in two--dimensional Potts
models}
\author{E. Carlon}
\address{Laboratoire de Physique des Mat\'eriaux, 
Universit\'e Henri Poincar\'e, Nancy I,\\
B. P. 239, F--54506 Vandoeuvre-l\`es-Nancy Cedex, France}
\author{F. Igl\'oi}
\address{Research Institute for Solid State Physics and Optics, H-1525 Budapest,
P.O.Box 49, Hungary\\Institute for Theoretical Physics, Szeged University
H--6720 Szeged, Hungary}
\author{W. Selke, F. Szalma}
\address{Institut f\"ur Theoretische Physik, Technische Hochschule\\
D--52056 Aachen, Germany}
 
\maketitle
 
\begin{abstract}
The interfacial adsorption $W$ at the first--order transition
in two--dimensional $q$--state Potts models is studied. In
particular, findings of Monte Carlo simulations and of density--matrix
renormalization group calculations, at $q= 16$, are consistent
with the analytic result, obtained in the Hamiltonian limit
at large values of $q$, that $W \propto t^{-1/3}$ on approach
to the bulk critical temperature $T_c$, $t=|T_c-T|/T_c$. In
addition, the numerical findings allow to estimate corrections to
scaling. Our study supports and specifies a previous conclusion by Bricmont
and Lebowitz based on
low--temperature expansions.\\

\end{abstract}

\vspace{5mm}

\vspace{1cm}

\noindent {\bf 1. Introduction}
 
The interface between two phases may become unstable
against the appearance of a third phase. This wetting phenomenon
has been studied in various circumstances, considering different
materials and geometries, both experimentally and
theoretically. \cite{diet}\\

Good candidates for modelling wetting are $q$-state Potts
models \cite{selke1,selke2}, where
the two phases correspond to distinct boundary
states, say, 1 and $q$, at opposite sides of the system. In that
case one observes an excess adsorption of the non--boundary
states at the interface between '1' rich and '$q$' rich domains.\\

In particular, in two dimensions, the interfacial adsorption, $W$, is found
to diverge on approach to the critical temperature $T_c$
like $W \propto t^{-\omega}$, where $t= |T_c -T|/T_c$. For
$q$= 3 and 4, the bulk transition is continuous and $\omega$ can be
expressed in terms of canonical bulk critical
exponents, $\omega= \nu -\beta$. \cite{selke1} For larger values
of $q$, the transition is of first order. On general, phenomenological
grounds \cite{lipo}, one may expect
then $\omega =1/3$, as observed in Monte Carlo (MC) simulations on the
two--dimensional Blume--Capel model. \cite{selke3} However, previous
simulations on the two--dimensional 20--state Potts model
suggested that $\omega$ may be significantly larger than
1/3, albeit a pronounced curvature in the
corresponding log--log plots for $W(t)$ was noticed indicating
that the asymptotic behavior may not have been reached. \cite{selke2} A
reasonable explanation of the numerical findings was then offered
by Bricmont and Lebowitz. \cite{bric} Based on low--temperature
expansions, the critical region is argued to be
extremely narrow in the Potts case, in contrast to
the Blume--Capel case. \cite{bric} However, a
numerical confirmation remained to be done.\\

Motivated by a recent intrigiung analysis of
the interfacial tension of the $q$--state Potts model in
two dimensions employing a field--theoretic representation
for configurations of the interface \cite{cardy} (extending
prior considerations \cite{selke1}), we decided to reconsider
the somewhat irritating problem on the value of $\omega$ for
$q > 4$. Indeed, advances in methods and
computer facilities allow now to explore numerically the
critical region more deeply than some years ago. In addition, we
dealt with the issue analytically by studying the Hamiltonian
limit of the Potts model for large values of $q$ exactly.\\

The outline of the article is as follows. In the next Section, the
model is defined, and the numerical
methods of our choice, MC simulations
and density--matrix renormalization group (DMRG)
calculations, are introduced. Then, we discuss the results obtained from
those methods. In Section 4, the analytical
findings in the Hamiltonian limit are presented. A summary concludes
the paper.\\

\noindent {\bf 2. Model and numerical methods}

The two--dimensional $q$--state Potts model is described
by the Hamiltonian

\begin{equation}
{\cal H} = -J~ {\sum\limits_{(ij),(i'j')}} \delta_{n_{ij},n_{i'j'}} 
\end{equation}

\noindent
summing over neighboring sites $(ij), (i'j')$ on a $L \times M$
rectangular lattice, with the Potts variable $n_{ij}= 1,2,...q$. In
the thermodynamic limit, $L, M \longrightarrow \infty$, the model
displays a bulk phase transition at the critical temperature
$k_BT_c/J= 1/\ln (\sqrt q +1)$ \cite{wu}, where $k_B$ is the
Boltzmann constant. The transition is of continuous type
at $q \le 4$, while it is of first order at larger values of $q$.\\

To introduce an interface, we add a column of $M$ sites on the
left side boundary, fixing the Potts variable
to be in the state $n_L= 1$, and
another column on the right side boundary with $n_R= q$. The Potts
variables on the top and at the bottom of the lattice may be connected by
periodic boundary conditions ('periodic case'). Alternatively, the
interface may be pinned at the lower and upper boundaries by inserting
there additional boundary rows with fixed states being '1' on the left half
and '$q$' on the right half (pinned case).\\

By examining typical Monte Carlo equilibrium configurations below $T_c$, it
is seen that an excess of non--boundary states is adsorbed at the
interface, as illustrated in Fig. 1. This wetting phenomenon may
be described quantitatively by the net interfacial adsorption $W$ per
unit length

\begin{equation}
W = 1/M~ {\sum\limits_{(ij),nb}} (< \delta_{nb,n_{ij}}>_{1:q}- <\delta_{nb,n_{ij}}>_{1:1} )
\end{equation}

\noindent
summing over all $L \times M$ sites in the inner part of the system; the
$<~>$ brackets refer to thermal averages; the index $nb$ denotes the
non--boundary states, $nb=2,...,q-1$; the subscripts 1:$q$ and 1:1 refer
to systems with corresponding fixed states at the boundaries, i.e. with
and without interface. The net adsorption $W$ is closely related
to the density profiles

\begin{equation}
n_{b_1:b_2}(s,i) = 1/M~ {\sum\limits_{j}} < \delta_{s,n_{ij}}>_{b_1:b_2}
\end{equation}

\noindent
describing the variation of the density of state $s$ by going from the
left side, fixed in state $b_1$, to the right side, fixed in
state $b_2$, of the lattice, summing over each column, $j=1,...M$, with
$i$ running from 1 to $L$. Obviously,  

\begin{equation}
W =  {\sum\limits_{i,nb}} ( n_{1:q}(nb,i)- n_{1:1}(nb,i))
\end{equation}

We computed profiles and net adsorption numerically, using
Monte Carlo techniques \cite{bin} and the density--matrix renormalization
group method.\cite{DMRGbook} In the MC simulations, the standard
single--variable flip algorithm was applied, for system sizes
$L \le 256$ and $M \le 2000$, studying mostly the periodic case, augmented
by a few runs for the pinned case. To obtain accurate equilibrium
data, we typically averaged over several
realizations (using different random numbers), each 
consisting of $10^6$ Monte Carlo steps per site, including 
at least $10^5$ steps
for equilibration. Error bars resulted from averaging over
the ensemble of realizations. Usually, we set $q= 16$, where the
bulk phase transition is strongly first order with a bulk correlation
length at $T_c$ of a few lattice spacings. \cite{carl2} To search for 
possible $q$--dependences, we also simulated models with
$q$= 20 and 40. Furthermore, because clusters, formed by neighboring
sites in the same state, seem to play an interesting role in
the wetting phenomenon, see Fig. 1, we also computed cluster
properties in the MC study.\\

In the DMRG method, one considers strip--like lattices, i.e.
$M \longrightarrow \infty$, while $L$ is finite. The
algorithm, introduced by White in 1992 \cite{White92} for 
the study of the low-lying spectrum of quantum spin chains, has been
extended in several directions. \cite{DMRGbook} In
the present study we follow Nishino's \cite{nishino1} formulation 
of the method adapted to treat classical two-dimensional systems,
where the approach is used for constructing iteratively approximate 
transfer matrices, starting from strips of small
width (say, $L= 8$) which can be also handled numerically exactly. 
At each DMRG step the strip width is increased, and the configurational 
space is truncated efficiently through the projection into smaller 
subspaces with the help of appropriate density matrices.

We do not need to describe details of the widely used DMRG procedure here: a
good introduction, together with recent developments, can be
found in Ref. \onlinecite{DMRGbook}. The 
DMRG technique has already been applied to the two--dimensional Potts
model, both at continuous, $q \leq 4$, \cite{nishino2,prb1} and
first--order transitions, $q > 4$ \cite{prb2}. In the
former case, bulk and surface critical exponents 
have been calculated with a high degree of 
accuracy \cite{prb1}. In the
present study we are interested in rather large values of $q$, where
the standard DMRG method is less suitable. An appropriate, powerful
variant, the "pseudo - spin" version, has 
been introduced in Ref. \onlinecite{prb2}. It enables one to
treat systems with special values of $q$, namely 
$q = p^2$, $p = 2, 3, \ldots$. We used that algorithm in the current
work. Most calculations were performed at $q = 16$, augmented
by some calculations at the
$q = 9$, for strips of widths up to $L = 80$. We kept up to $m = 60$
states per block with a typical truncation error of $\epsilon
\approx 10^{-6}$. From the dominant eigenvector of the transfer matrix the
profiles \cite{prb1} and the interfacial adsorption $W$ were
calculated.\\

\noindent {\bf 3. Numerical results}

The crucial quantity, which we
computed numerically, is the net adsorption $W$ as a function
of the system size, $L, M$, and temperature $t$, $W(L,M,t)$. The 
main aim is to analyse its critical behavior
as $t \longrightarrow 0$ or $T \longrightarrow T_c$, in the 
thermodynamic limit $L, M \longrightarrow \infty$. For that, one
may extrapolate the numerical
data for $W$ to that limit, or one may
study systems for which finite--size effects can be neglected. We 
also tried to perform a finite--size scaling analysis on
$W(L,M,t)$ \cite{selke3}, see below.\\

In any event, accurate data are needed. Their quality
can be conveniently tested by comparing results obtained from the
MC simulations and the DMRG calculations, as illustrated in 
Figs. 2 and 3.\\

In Fig. 2, profiles, Eq. (3), are shown for the 16--state Potts model
with and without interface, demonstrating again the excess
adsorption of non--boundary states at the interface. In Fig. 3, the
increase of the net
adsorption $W$ with increasing width $L$ of the Potts model, at various
temperatures, is displayed. In both figures, the finite--size effect
arising from the length $M$ of the system can be disregarded. In
fact, $M$ is infinite in the DMRG approach. In the
simulations (periodic case), $M$ was checked to be
sufficiently large so that $W$
approached closely $W(L,M=\infty ,t)$, with the characteristic
crossover value depending, of course, on the width $L$ and the
distance from criticality $t= |T_c- T|/T_c$.\\

As exemplified in Figs. 2 and 3, data from both methods do, indeed,
agree nicely, being obviously accurate and reliable. Slight
systematic deviations seem to become significant only for
quite wide systems, say, $L > 60$. A reasonable extrapolation of our
data for $W$ to the thermodynamic limit is
feasible for temperatures $T \le 0.998 T_c$, see Fig. 3. Closer
to $T_c$, both numerical methods would require much larger
system sizes demanding extremely large storage and/or computing
time.\\

In the thermodynamic limit, one expects $W \propto t^{- \omega}$ as
$t \longrightarrow 0$. To monitor the approach to the 
asymptotic behavior, one may consider the effective exponent

\begin{equation}
\omega _{eff}(t) = -d \ln W/d \ln t
\end{equation}

\noindent
with $\omega = \omega_{eff}(t=0)$. In Fig. 4, numerical estimates
of the effective exponent are shown. At $t \ge 0.02$, we
determined $\omega _{eff}$ from MC data for systems being
large enough to disregard finite--size effects. Actually, finite--size
effects are much stronger in the pinned case than in the
periodic case, and only MC data for the latter case are included, with
$L= 64$ and $M \ge 400$. At $t \le 0.01$, estimates
are based on extrapolating MC and DMRG results for $W$ to the
thermodynamic limit, see Fig. 3. Note that the net adsorption is
calculated at discrete temperatures, $t_i$, and the effective
exponent may be approximated by $\omega_{eff}(t) =
-\ln (W(t_i)/W(t_{i+1}))/ \ln (t_i/t_{i+1})$, with $t$=
 $\sqrt{(t_i t_{i+1})}$, and $t_{i+1} < t_i$. Error bars in
Fig. 4 stem from a proliferation of the error in the net
adsorption $W$. Evidently, $\omega_{eff}$ depends strongly on the
distance from criticality $t$. For instance, in the 
range $0.1 > t > 0.002$, it changes from about 1.2 to about 0.5 when
moving towards $T_c$. Accordingly, an average critical exponent
in that interval would be supposedly significantly larger
than the true asymptotic critical exponent
$\omega$, as observed before. \cite{selke2} Presuming $\omega =1/3$, see
Refs. 4 and 6 as well as Section 4, it follows from
Fig. 4 that the asymptotic regime is
very narrow, and corrections to scaling are quite pronounced. To
quantify these corrections, one may postulate the standard ansatz 

\begin{equation}
 W(t) = W_0 t^{-1/3} ( 1 + a t^x + b t^{2x} + ...)
\end{equation}

\noindent
The coefficients can be calculated from chi--square fits to 
the numerical data near $T_c$. Using, for instance, a fit to the points
in the interval $0.02 \le t \le 0.1$ (shown in Fig. 4), leads to 
a net adsorption $W$
reproducing very well the numerical findings both closer to $T_c$
and further away from $T_c$. Eventually, systematic deviations
show up when further lowering the temperature, reflecting the need for
additional correction terms in Eq. (6) in that region, see Fig. 4. The 
exponent $x$ characterising the corrections to scaling
is found to be rather small, $x= 0.14 \pm 0.06$. Because of its
smallness, we included the leading, with the 
exponent $x$, and subleading, $2x$, terms
in the ansatz (6). The error bars
arise from using a variety of plausible fitting intervals
and points.\\

From MC simulations of Potts models with larger
number of states, $q= 20$
and 40, one may conclude that the corrections to scaling, at
$t > 0.01$, are rather
insensitive to the concrete value of $q$.\\

From general considerations \cite{lipo,selke3}, one
expects two diverging lengths at the interface, in the direction
parallel to the interface, $\xi _{\parallel}$, and perpendicular
to it, $\xi _{\perp}$, with
$\xi _{\perp} \propto t^{-\omega}$ and 
$\xi _{\parallel} \propto t^{- 2\omega}$, as $t \longrightarrow 0$, see 
also Section 4. For strip--like systems, $M \longrightarrow \infty$, the
following finite--size scaling expressions
can be then motivated \cite{selke3}

\begin{equation}
 W(L, M=\infty, t) = t^{-\omega} w_1(L t^ \omega)
\end{equation}

\noindent
in the limit of $L t^\omega >> 1$, and

\begin{equation}
 W(L, M=\infty, t=0) \propto L
\end{equation}

\noindent
for $L >> 1$. Indeed, the DMRG results indicate that Eq. (8) seems
to be satisfied rather well already for strips of moderate
width, say, $L \ge 16$, see also Ref. 3. On the other hand, the numerical
data do not suffice to establish the scaling form (7) with 
$\omega= 1/3$. In fact, a 'reasonable' collapse of our data on an
apparent scaling function $w_1$ might be achieved with
a somewhat larger value, $\omega \approx 1/2$. Careful
inspection, however, reveals that systematic deviations from a
unique scaling curve $w_1$ set in for large
arguments $L t ^\omega$. Thence, larger systems close to $T_c$ had to
be studied (which are, at present, out of reach, because
of limitations in storage and computing
time). Indeed, $\omega \approx 1/2$ is merely
an upper bound of the true value of $\omega$. This
observation corroborates the above mentioned finding on  
$\omega_{eff}$: One has to include corrections to scaling to
demonstrate consistency of the numerical data with the
theoretically expected asymptotic behavior, due
to the narrowness of the critical region.\\

In closing the Section on the MC and DMRG results, we
remark that the thermally averaged largest cluster of
non--boundary states, as computed in the simulations, seems
to diverge by approaching $T_c$ from below. A detailed
analysis would be desirable, but it is beyond the
scope of the present study.\\

\noindent {\bf 4. The Hamiltonian limit for large values of $q$}

\newcommand{\bc}{\begin{center}}
\newcommand{\ec}{\end{center}}
\newcommand{\be}{\begin{equation}}
\newcommand{\ee}{\end{equation}}
\newcommand{\beqn}{\begin{eqnarray}}
\newcommand{\eeqn}{\end{eqnarray}}

In the following we consider the Hamiltonian limit of the Potts model
with strong vertical and weak horizontal
couplings\cite{kogut}. The transfer matrix in the vertical
direction has the form
${\cal T}=\exp(- \hat{ H})$, with the one--dimensional
Hamiltonian $\hat{H}$ \cite{solyom}
\be
\hat{ H}=-\sum_{i=1}^{L-1} \delta_{n_i,n_{i+1}} - h \sum_{i=1}^L \sum_{k=1}
^{q-1} M_i^k\;.
\label{hamilton}
\ee
$n_i$ is the Potts variable on site $i$ and $M_i^k$ denotes the  
flip operation $M_i^k |n_i \rangle =|n_i+k, {\rm mod}~q
\rangle$. The strength of the transverse field, $h$, at the transition
point is $h_c=1/q$. Quantities
of physical interest are derived from the ground state, $|\Psi_0
\rangle$, and from the energies of the ground state and
the first excited state, $E_0$ and $E_1$, of (9).\\

The Hamiltonian limit of the Potts model has been treated
recently \cite{prb2} for free boundary
conditions. In that case, the solution has a
remarkably simple form in the vicinity of the transition point for large
values of $q$. Repeating the same type of considerations for models
with an interface, fixing the variables at the boundaries in the
states '1' and '$q$' (1:$q$) (as before), one finds that the
ground state sector of
the Hamiltonian (9) is spanned by the vectors
\be
|\psi_{i,j} \rangle=|11 \dots 1n_b n_b \dots n_b qq \dots q\rangle \;,
\label{psi}
\ee
where a non-boundary state is given by
$|n_b \rangle=1/\sqrt{q}(|2 \rangle +|3 \rangle +\dots+|q-1 \rangle)$, and the
positions of the domain walls
separating the boundary and non-boundary states are denoted by
$i(= 1,2,\dots,L$) and $j=( i,i+1,\dots,L$). The diagonal matrix-elements of
these states, $\langle \psi_{i,j}|\hat{ H}| \psi_{i,j} \rangle=-L-(j-i)t$,
with $t(= hq-1$, $|t|\ll 1$) being the distance from the critical point, are
smaller by an amount of $O(1)$ compared to any other states,
like those containing boundary states
in the domain of non-boundary ($nb$) states. The Hamiltonian
in the ground state sector, spanned by the vectors (\ref{psi}), can be
written as
\be
\hat{ H}_g=-(j-i)t-h\sqrt{q}[(a^+ + a^-) + (b^+ + b^-)]\;,
\label{hamiltong}
\ee
up to a constant; the operators, $a^{\pm}$ and $b^{\pm}$, which move the
positions of the domain walls in (\ref{psi}), are defined as
\beqn
a^{\pm} | \psi_{i,j} \rangle &=&|\psi_{i\pm 1,j} \rangle,~~1<i<j;~~~
a^+ |\psi_{i,i} \rangle=a^- |\psi_{1,j} \rangle=0\cr
b^{\pm} |\psi_{i,j} \rangle &=&|\psi_{i,j \pm 1} \rangle ,~~i<j<L;~~~
b^- |\psi_{i,i} \rangle =b^+ |\psi_{i,L} \rangle =0\;.
\label{ab}
\eeqn

In the continuum limit, when $L\gg 1$, $i \gg 1$ and $j \gg 1$, but
$x=i/L=O(1)$ and $y=j/L=O(1)$, the Hamiltonian (\ref{hamiltong})
can be written in the form of a differential operator
\be
\hat{ H}_g \psi(x,y)=-\left[{h \sqrt{q} \over L^2} \left({\partial^2 \over
\partial x^2} + {\partial^2 \over \partial y^2}\right)+tL(y-x)\right] \psi(x,y)
=E \psi(x,y)\;,
\label{diffeq}
\ee
with the boundary condition $0\le x \le y \le 1$.

{\it At the critical point}, $t=0$, the solution of the eigenvalue
problem (\ref{diffeq}) reads
\be
\psi (x,y)=2 [ \sin(\pi k_1 x) \sin(\pi k_2 y)- \sin(\pi k_2 x)
\sin(\pi k_1 y) ]\;,
\label{psit0}
\ee
with $k_1=1,2,\dots$ and $k_2=k_1+1,k_1+2,\dots$. For the ground
state, $\psi_0(x,y)$, one has
$k_1=1$ and $k_2=2$, whereas for the first excited state $k_1=1$ and
$k_2=3$. Thus the energy gap is $\Delta E=E_1-E_0=5 \pi^2 h \sqrt{q}
L^{-2}$, and the correlation length parallel to the
interface behaves as $\xi_{\parallel} \sim (\Delta E)^{-1} \sim L^2$.
Since the correlation length perpendicular to the interface is, at the
critical point, limited
by the width of the system, $\xi_{\perp} \sim L$, one arrives at
$\xi_{\parallel} \sim \xi_{\perp}^2$, in agreement with the form
mentioned above.

The density profiles satisfy the relations
$n_{1:q}(1,x)=n_{1:q}(q,1-x)$ and
$n_{1:q}(x)= {\sum\limits_{nb}} n_{1:q}(nb,x)=
 1-n_{1:q}(1,x)-n_{1:q}(q,x)$. From the ground state, one obtains
\beqn
n_{1:q}(1,x)&=&\int_x^1 \rm{d} x' \int_{x'}^1 {\rm d} y \left[ \psi_0(x,y) \right]^2\nonumber \\[0.2cm] &=&
{4 \over \pi^2} \left\{ \left[{\pi \over 2}(1-x) + {1 \over 4} \sin 2 \pi x \right]
\left[{\pi \over 2}(1-x) + {1 \over 8} \sin 4 \pi x \right]
-{4 \over 9} \sin^6 \pi x \right\}\;.
\label{1profil}
\eeqn
For small $x$, one finds $n_{1:q}(1,x)=1- (10 \pi^2/ 3) x^3 + O(x^6)$,
whereas for $x$ close to one, the profile behaves
like $n_{1:q}(1,x) \sim (1-x)^{10}$. The profile of non-boundary
states is symmetric and its
maximal value is given by $n_{1:q}(1/2)= 1/2 + 32/(9 \pi^2)=0.86025$.

For the $(1:1)$ boundary condition, the profile
of non-boundary states tends to
zero for large values of $q$. Therefore the interfacial adsorption $W$, Eq.
(2), at the transition point may be approximated by
\be
{W \over L}= \int_0^1 n_{1:q}(x) {\rm d} x ={1 \over 3} + {35 \over 72 \pi^2}=0.3826\;.
\label{Wcrit}
\ee
Thus, at the transition point, $W$ is, indeed, proportional to $L$, see
Eq. (8). Note that the prefactor seems to depend on
$q$, being, at $q= 16$, about 0.3, according to the DMRG calculations.\\

{\it Below the critical point}, $t \ne 0$, we consider the eigenvalue equation
(\ref{diffeq}) in terms of the new variables $x_+=(y+x)/\sqrt{2}$ and
$x_-=(y-x)/\sqrt{2}$. Then
\be
-\left[{h \sqrt{q} \ 2 \over L^2} \left({\partial^2 \over
\partial x_+^2} + {\partial^2 \over \partial x_-^2}\right)+tL\sqrt{2}x_-\right] \psi(x_+,x_-)
=E \psi(x_+,x_-)\;,
\label{diffeq+-}
\ee
with the boundary condition $0\le x_- \le x_+ \le 1/\sqrt{2}$ and
$0 \le x_-\le \sqrt{2}-x_+ \le 1/\sqrt{2}$. Now the
eigenfunction $\psi$ can be written as
 $\psi(x_+,x_-)=\phi_+(x_+) \phi_-(x_-)$.
 $\phi_+(x_+)$ satisfies the free-particle equation
$-{\rm d}^2 \phi_+/{\rm d} x_+^2= L^2E_+/(2 h \sqrt{q}) \phi_+$; 
 $\phi_-(x_-)$ is the solution of the Schr\"odinger equation of
a particle in a linear potential
\be
-\left[{h \sqrt{q} \ 2 \over L^2} {\partial^2 \over
\partial x_-^2} + +tL\sqrt{2}x_-\right] \phi_-(x_-)
=E_- \phi(x_-)\;.
\label{diffeq-}
\ee
Equation (\ref{diffeq-}) leads to bound states, and the energy
scale (both for the ground state and the excited states) is set by
$\epsilon \sim t^{2/3}$. Hence the temperature dependence of the parallel
correlation length is given by $\xi_{\parallel}\sim (\Delta E)^{-1} \sim
t^{-2/3}$, in accordance
with the phenomenological considerations\cite{lipo}. On the other hand, the
extent of the bound states sets the length
scale $\xi_{\perp} \sim t^{-1/3}$,
which is then proportional to the interfacial adsorption, W. Thence, in
the Hamiltonian limit for large values of $q$, one has
$\omega=1/3$.

\noindent {\bf 5. Summary}

In this article, critical interfacial properties of
two--dimensional $q$--state Potts models at the bulk first--order
phase transition have been studied. We applied two
numerical methods, Monte Carlo simulations and the
density--matrix renormalization group approach, mainly
at $q= 16$. Furthermore, we
considered analytically the model in its Hamiltonian limit at
large values of $q$.\\

The different methods lead to a consistent description of the
critical behavior. The interfacial adsorption $W$ diverges
on approach to the phase transition temperature
as $W \propto t^{-\omega}$, $\omega =1/3$, with a very narrow
asymptotic region. The strong corrections to scaling are characterised
by a small exponent, $x= 0.14 \pm 0.06$, which seems to depend (if
at all) only weakly on the number of Potts states, $q$. At
the critical point, $W$ diverges linearly with the width of
the system (being indefinitely long in the direction parallel
to the interface). The proportionality factor has been
calculated in the Hamiltonian limit.\\

The value of the critical exponent $\omega$, $\omega =1/3$, is
typical for wetting phenomena at bulk transitions of first order
in two dimensions. It follows from
general, phenomenological considerations, based
on an effective interface Hamiltonian \cite{lipo}, as well
as from calculations on various microscopic multi--state models, such
as Potts and the Blume--Capel models. The strong corrections to
scaling and the narrowness of the critical region are, on the 
other hand, features which are specific for Potts models. They 
have been predicted before \cite{bric} by using
low--temperature arguments, and they have been quantified
in this study.\\

\noindent {\bf Acknowledgements}
  
F. Sz. would like to thank
the DAAD for a scholarship
enabling his visit at the RWTH Aachen.
The work of F. I. and F. Sz. has been supported by the National Research
Fund under grant No. OTKA TO23642, TO25139, F/7/026004, M 028418 and by the
Ministery of Education under grant No. FKFP 0765/1997. Useful discussions with
L. Turban are gratefully acknowledged.\\

\begin{figure}
\caption{Typical Monte Carlo equilibrium configuration of the
two--dimensional $16$--state Potts model
 with an interface, at $T= 0.99T_c$. The '1'
domain is on the left hand side, and the '16' domain on the
right hand side. Shown are clusters of distinct states. A MC system
of size $L$ = 60
and $M$ = 60, periodic case, was simulated, but only a part is depicted.}
\label{fig1}
\end{figure}
 
\begin{figure}
\caption{Profiles, summed over the non--boundary ($nb$)
 states, $n_{1:b}(i) =  {\sum\limits_{nb}}  n_{1:b}(nb,i)$ for
systems with interface, $b=16$, and
without interface, $b=1$, at $T= 0.99 T_c$ and $L=24$, as
obtained from the DMRG method (open circles) and MC
simulations (periodic case with $M=400$; full diamonds).} 
\label{fig2}
\end{figure}
 
\begin{figure}
\caption{Net adsorption, $W$, vs. inverse width, $1/L$, of the
lattice at (from bottom to top) $t= 0.01, 0.005, 0.002, 0.001$, and
0.0005, depicting DMRG (open circles) and MC (periodic 
case with $M \le 1000$; full
diamonds) data.}
\label{fig3}
\end{figure}
 
\begin{figure}
\caption{Temperature, $t$, dependence of the effective exponent
of the net adsorption $W$,  $\omega_{eff}(t)$, for
numerical data 'free of finite--size effects', see text. The 
dashed curve corresponds to the fit to Eq. (6), with $W_0= 3.280$,
 $a= -1.977$,
 $b= 0.939$, and $x= 0.165$, quantifying the corrections to scaling.}
\label{fig4}
\end{figure}

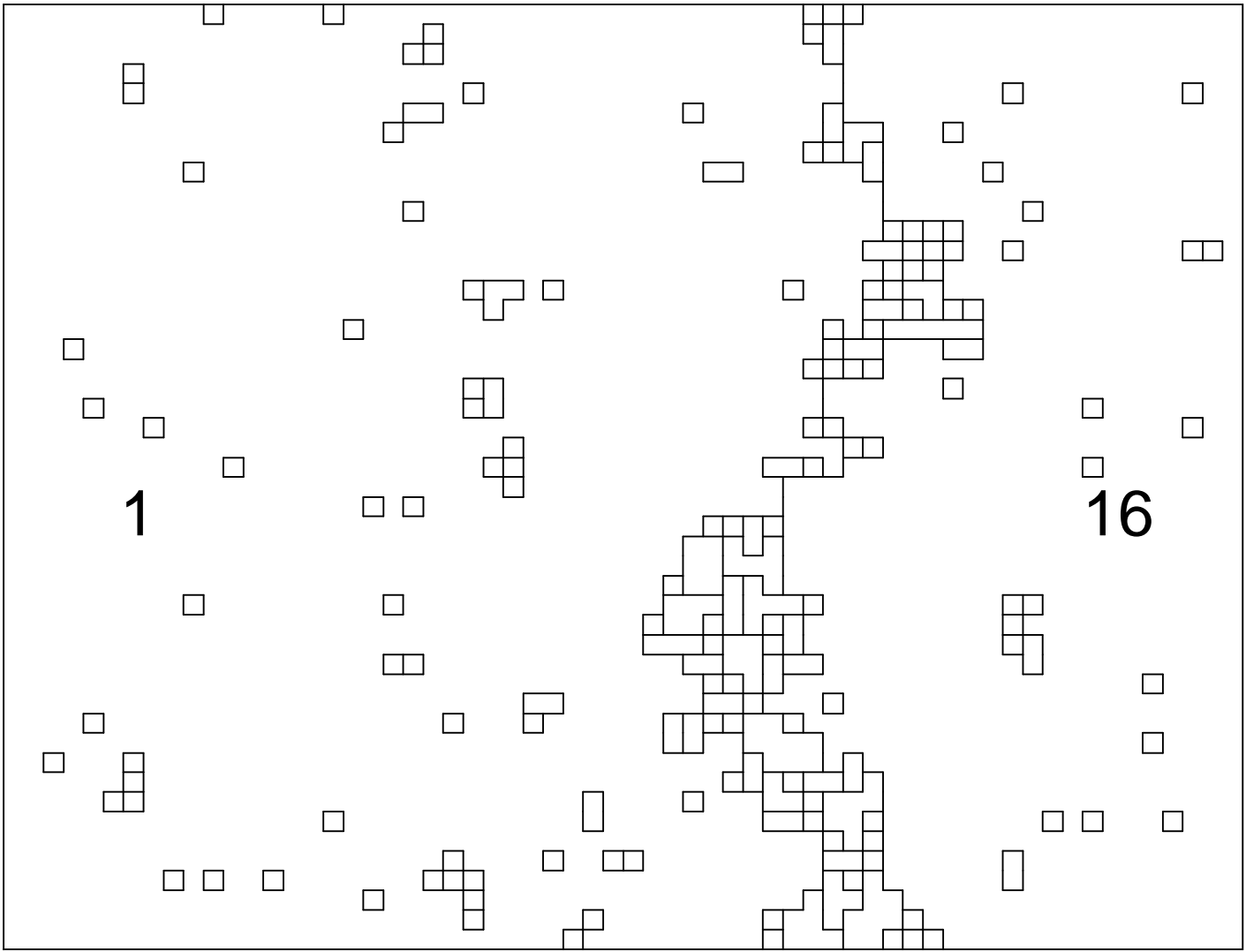
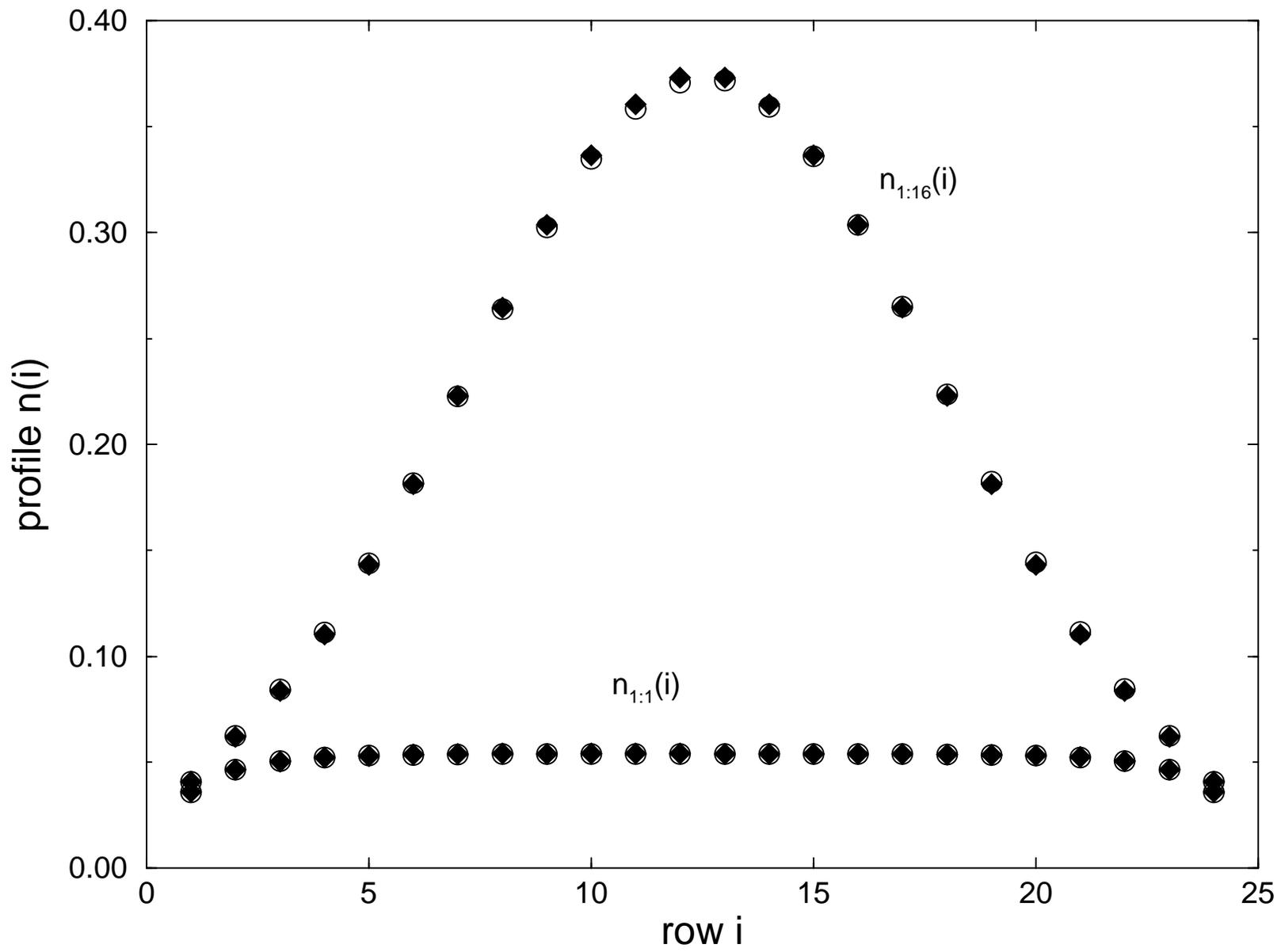
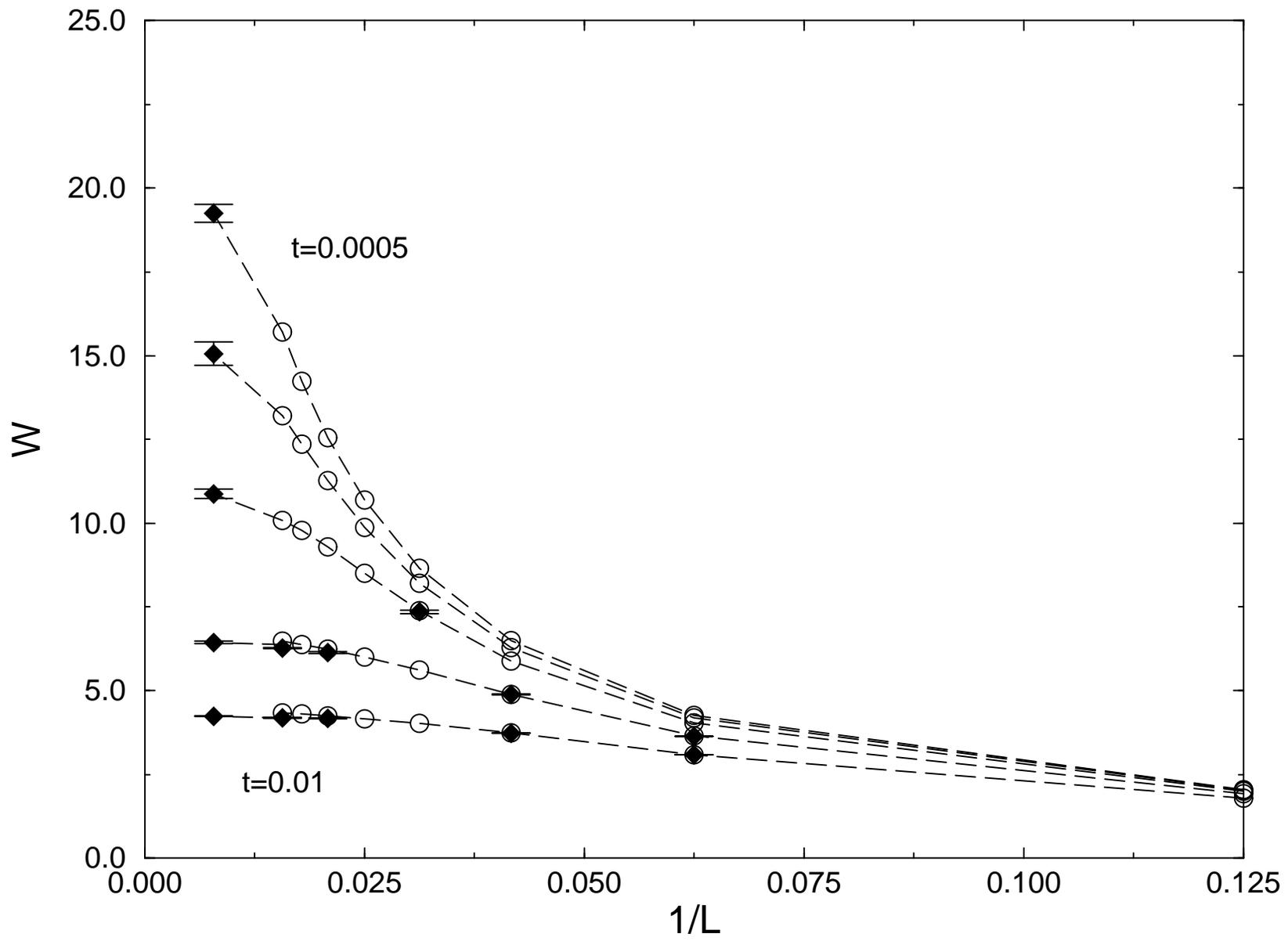
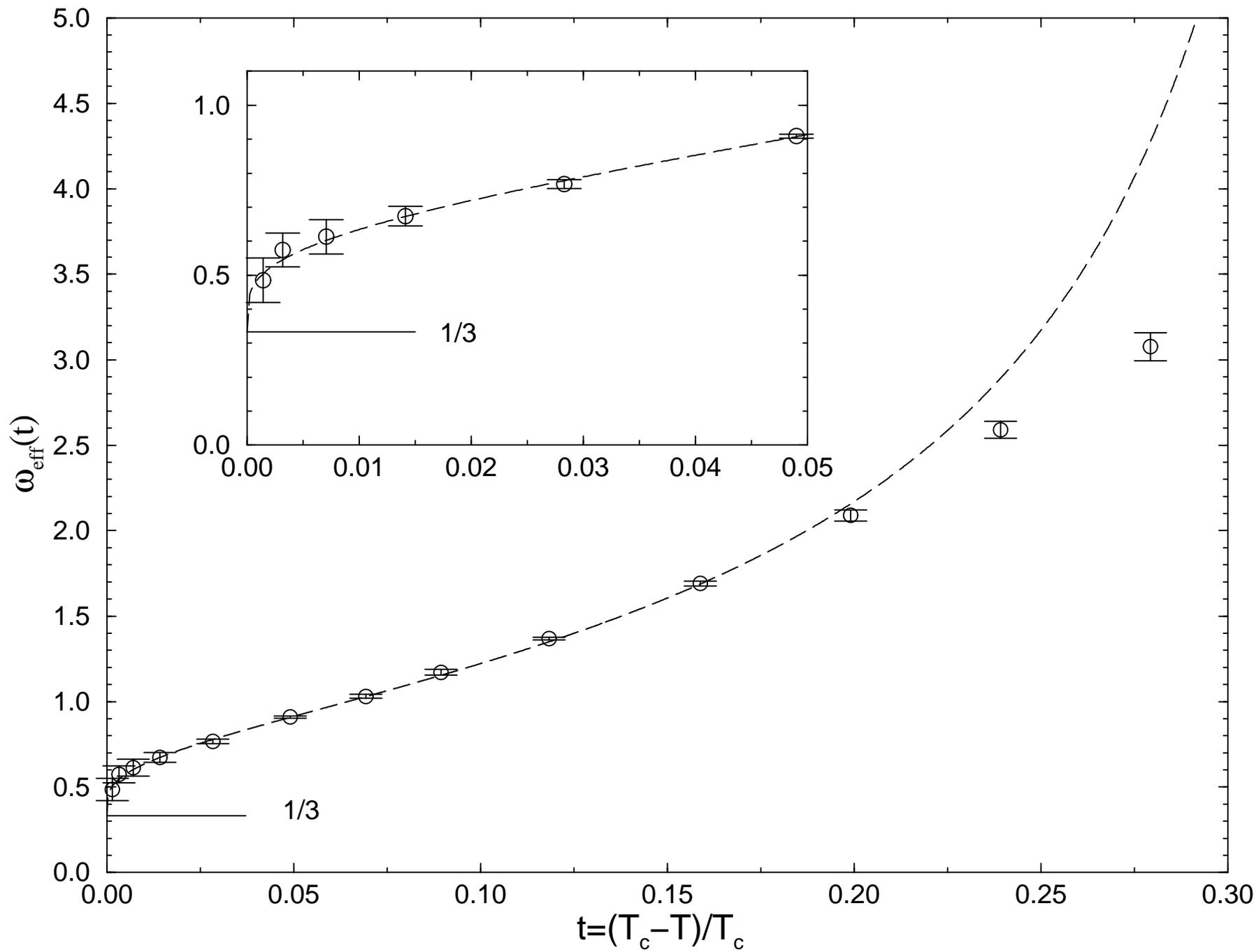
\end{document}